\documentclass[prb,superscriptaddress,reprint,amssymb,aps,floatfix,showkeys]{revtex4-1}
\usepackage{graphicx}
\usepackage{amsmath}
\usepackage{dcolumn}
\usepackage{bm}
\usepackage{bbold}
\usepackage{color}
\usepackage{amsfonts}
\usepackage{amssymb}
\usepackage{mathrsfs}
\usepackage{braket}
\usepackage{color}

\renewcommand\Re{\operatorname{\mathfrak{Re}}}
\renewcommand\Im{\operatorname{\mathfrak{Im}}}

\begin{document} 

\title{Controlling a whispering gallery doublet mode avoided frequency crossing: Strong coupling between photon bosonic and spin degrees of freedom}

\author{Maxim Goryachev}
\affiliation{ARC Centre of Excellence for Engineered Quantum Systems, University of Western Australia, 35 Stirling Highway, Crawley WA 6009, Australia}

\author{Warrick G. Farr}
\affiliation{ARC Centre of Excellence for Engineered Quantum Systems, University of Western Australia, 35 Stirling Highway, Crawley WA 6009, Australia}

\author{Daniel L. Creedon}
\affiliation{ARC Centre of Excellence for Engineered Quantum Systems, University of Western Australia, 35 Stirling Highway, Crawley WA 6009, Australia}

\author{Michael E. Tobar}
\email{michael.tobar@uwa.edu.au}
\affiliation{ARC Centre of Excellence for Engineered Quantum Systems, University of Western Australia, 35 Stirling Highway, Crawley WA 6009, Australia}

\date{\today}


\begin{abstract}

A combination of electron spin interactions in a magnetic field allows us to control the resonance frequencies of a high-Q Whispering Gallery (WG) cavity mode doublet, resulting in precise measurements of an avoided crossing between the two modes comprising the doublet. We show that the resonant photons effectively behave as spin--$\frac{1}{2}$ particles and that the physical origins of the doublet phenomenon arise from an energy splitting between the states of photon spin angular momentum. The exclusive role of the photon spin in splitting the mode frequency is emphasized, and we demonstrate that the gyrotropic and anisotropic properties of the crystalline media supporting the WG mode lead to strong coupling between the bosonic and spin degrees of freedom of cavity photons. Despite the demonstrated similarities with Jaynes-Cummings type systems, the mode doublet system exhibits a significant difference due to its linearity. Unlike traditional experiments dealing with interactions between fields and matter, here the crystalline medium plays a role of macroscopic symmetry breaking, assisting in the strong coupling between these photon degrees of freedom. Such a regime is demonstrated experimentally with a method to effectively control the photon spin state. Our experiments demonstrate for the first time, controllable time-reversal symmetry breaking in a high-$Q$ cavity. 
\end{abstract}

\maketitle

\section{Two Approaches to Field-Matter Interaction}

The interaction between light and matter is one of the most important topics in modern science and technology, and particularly in the field of quantum mechanics. The applications of such research are many and varied, including spectroscopy, quantum information and computing, ultra-stable clocks, lasers, and fundamental investigations of  quantum phenomena. Light-matter interactions are investigated at almost all frequencies of the electromagnetic spectrum, spanning many orders of magnitude from ultraviolet to radio frequencies. Generally speaking, in all of these experiments the bosonic degree of freedom of photons is always coupled to quantum states of matter in the form of atoms, ions, molecules or plasmas. Photons are chosen due to well established experimental techniques and technologies that make experimentation very easy, whilst the aforementioned states of matter are chosen as they are known for exhibiting quantum mechanical phenomena due to their proximity to the Plank scale. This approach has been enormously successful over the past few decades in demonstrating the quantum mechanical properties of nature.

One of the most common types of experiment involves coupling electromagnetic radiation to a spin$-\frac{1}{2}$ particle such as an electron. Since the famous Stern-Gerlach experiment of 1922, experimental setups involving spin$-\frac{1}{2}$ particles have gained significant attention in physics. In particular, systems well described by the Jaynes-Cummings model\cite{JC} representing the coupling between a two-level system (TLS) with a quantised mode of an electromagnetic field, form the basis of a large area of research known as Cavity or Circuit Quantum Electrodynamics (QED)\cite{walther}. Quantum strong coupling regimes in particular\cite{aoki,hann} have attracted special interest, having demonstrated entangled states necessary for different types of quantum computing schemes. In the microwave frequency range, strong coupling has been demonstrated for a range of systems including ensembles of paramagnetic impurity ions in dielectric crystals\cite{pavel1}, nitrogen-vacancy (NV) centres in diamond\cite{NV1, NV2}, cold polar molecules coupled to superconducting cavities\cite{rabl}, NV centres coupled to qubits\cite{NV3}, and ensembles of ultra-cold atoms on a chip\cite{ritsch}. In all of these experiments, a special role is attributed to matter which reveals its microscopic quantum properties, particularly where spin systems play the role of two-level states.  The number of such spin systems suitable for QED experiments in the microwave range is very limited, and thus the search for other potential systems for QED purposes is important.

Another approach to field-matter interaction places only secondary importance on the matter. Here, its role is limited to macroscopic phenomena leading to the symmetry breaking. As a result of such an interaction with matter, a cavity photon bosonic degree of freedom is coupled to the spin angular momentum of the same photon. Despite the fact that a photon is a spin-1 particle, the photon spin degree of freedom is effectively seen as a TLS due to the zero rest mass of a photon. In such a way, the situation of a cavity mode coupling to a TLS is achieved even though no atom-like structure of matter is needed. However, this system cannot be classified as a Jaynes-Cummings interaction because it lacks any nonlinear properties naturally exhibited by such systems.

In this work, we present the first experimental and theoretical demonstration of a field-matter interaction experiment performed using the spin angular momentum of a photon\cite{oppenheimer,hiroshi}. Both the bosonic degree of freedom and the TLS originate in the nature of a cavity photon. The only role of matter in the experiment is to break reflection symmetry in such a way that both of these degrees of freedom of a single particle are coupled. For this purpose, a Whispering Gallery mode cavity (WGC) of cylindrical geometry\cite{karim1,kipp1} is used, rather than the more conventional linear Fabry-P\'{e}rot cavity. We demonstrate interaction in the strong coupling regime between a microwave photon spin (effectively acting as a TLS) and bosonic degrees of freedom. Although the quantum polarization state of photons has been studied previously in the optical domain\cite{hiroshi, karassiov, raymer,hostein}, no cavity experiment has been conducted and analysed yet. Unlike other works on WGC\cite{weiss, gorod,mazzei}, we emphasize importance of the polarization states of the photons, with an efficient method to control them.This gives an insight into the physical origins of the effects observed in many similar systems due to the spin degree of freedom of a photon.
 
\section{Experimentally Controlling a Photon Spin State}

In order to observe the energy splitting due to photon spin angular momentum, a few conditions must be satisfied. First, the cavity needs to be represented in a circular geometry. This requirement is imposed because a conventional linear cavity has an explicitly broken symmetry in reflection due to two boundary conditions. To excite a circular cavity requires only one electrode, and thus only one boundary condition is explicitly set. This implies independence of the clockwise and counter-clockwise propagating waves in an ideal cavity.  Secondly, the energy splitting must be greater than the resonance linewidth. In the opposite case, the relations between the mode doublets cannot be observed. Finally, an effective means is required to tune the cavity properties in order to change the coupling between cavity photons with different spin states. All these requirements can be achieved in ultra-low loss dielectric cylindrical cavities such as cryogenically cooled sapphire whispering gallery mode resonators\cite{karim1}. In such a system, the magnetic properties of the medium can be effectively manipulated by an external DC magnetic field $B_{\text{ext}}$ through intrinsic crystal impurity ions\cite{karim2}. Indeed, impurity ions such as Fe$^{3+}$ in sapphire crystal at low temperatures\cite{KorPro} change the permeability tensor $\hat\mu(B_{\text{ext}})$ making it gyrotropic \cite{karim2,gurevich}. The anisotropy of the permittivity tensor is due to the crystal structure of the sapphire resonator, as well as the presence of back-scatterers in the lattice\cite{weiss, gorod,giordano}. 

\begin{figure}[t!]
\centering
\includegraphics[width=89mm]{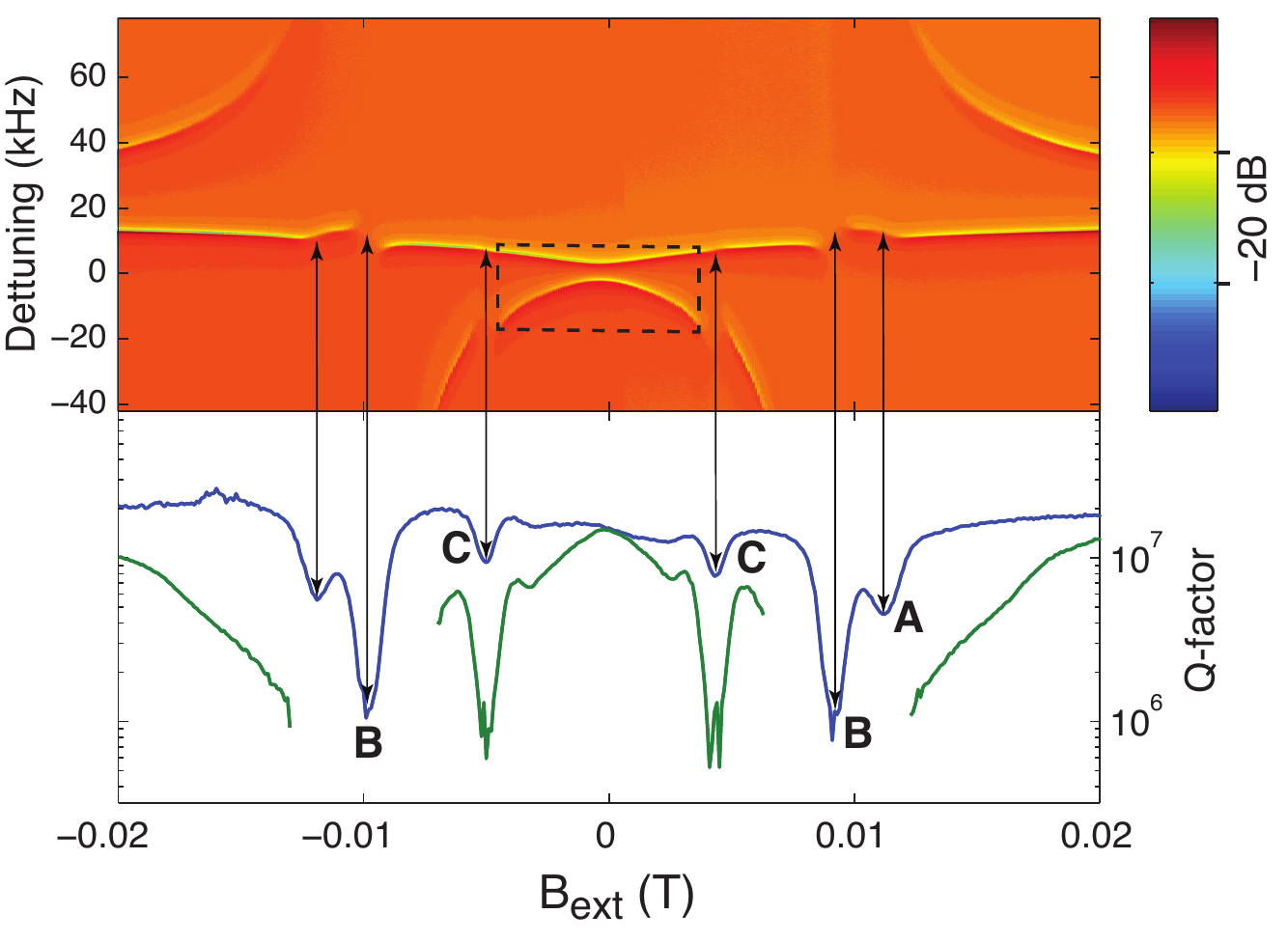}
\caption{\label{thing} (Color online) Density plot showing the dependence of the WG mode doublet transmission coefficient on the external applied magnetic field. The detuning frequency is calculated from $f_c=11.77355$~GHz. The electron spin resonance transitions labelled on the plot are the Cr$^{3+}$ dipole transition (\textbf{A}), Fe$^{3+}$ dipole transition (\textbf{B}), and Fe$^{3+}$ quadruple transition (\textbf{C}). The bottom section of the plot shows the $Q$-factor of the WG mode for corresponding states of photon spin angular momentum. }
\end{figure}

\begin{figure}[t!]
\centering
\includegraphics[width=89mm]{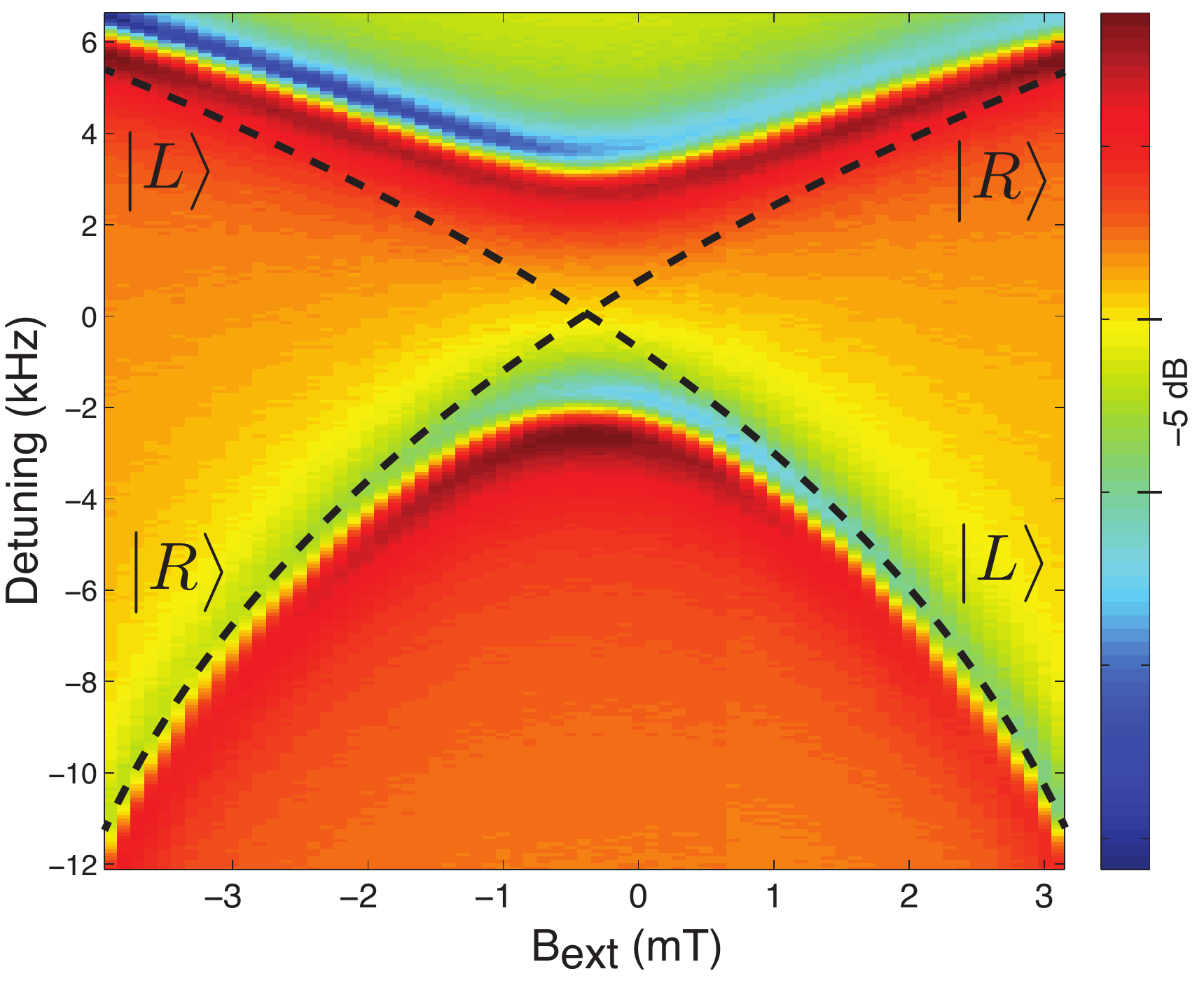}
\caption{\label{zoom} (Color online) Density plot showing the dependence of the WG mode transmission coefficient on the external applied magnetic field, and detuning from $f_c$. The dashed lines correspond to the eigen-energies of the bare states, and the transmission extrema correspond to ``dressed" states.  }
\end{figure}

To experimentally demonstrate coupling between the spin and bosonic degrees of freedom of a photon, a sapphire WGC was cooled to approximately $110$~mK. The sapphire is a cylinder of 50 mm diameter $\times$ 30mm height, characterised using a transmission method with two loop antennae in a setup described in detail elsewhere \cite{karim2}. The probing signal is attenuated to reach the level of $-60$~dBm power incident upon the crystal, and the cavity output signal is amplified by a low noise cryogenic amplifier. The cavity and the amplifier are both isolated at  $110$~mK. To modify the effective permeability sensor due to intrinsic impurity ions included at a parts-per-billion level within the crystal, an external DC magnetic field is applied to the sapphire cavity along its cylindrical $z$-axis. For the best magnetic field sensitivity $\hat\mu(B_{\text{ext}})$, the WG mode to be excited is chosen in the vicinity of an Electron Spin Resonance (ESR) of the residual Fe$^{3+}$ ions in the crystal (\hbox{$\nu_{\text{ESR}} =12.04$ GHz}). Fig.~\ref{thing} shows the response of the cavity near the Fe$^{3+}$ and Cr$^{3+}$ spin resonances, demonstrating the existence of two eigensolutions. These eigensolutions correspond to the $\Ket{R}$ and $\Ket{L}$ states of the photon spin angular momentum. The boxed region is shown magnified in Fig.~\ref{zoom}. This figure demonstrates an avoided crossing between two states of defined circular polarisation. The magnetic field effectively ``inverts" the medium, which manifests as a mirroring of these polarisations. The minimal splitting at about $-0.5$~mT is attributed to the electrical properties of the crystal, i.e. the permittivity tensor $\hat{\varepsilon}$. This parameter is set by the crystalline medium and cannot be varied, thus determining the minimal splitting.

\begin{figure}[t!]
\centering
\includegraphics[width=3.4in]{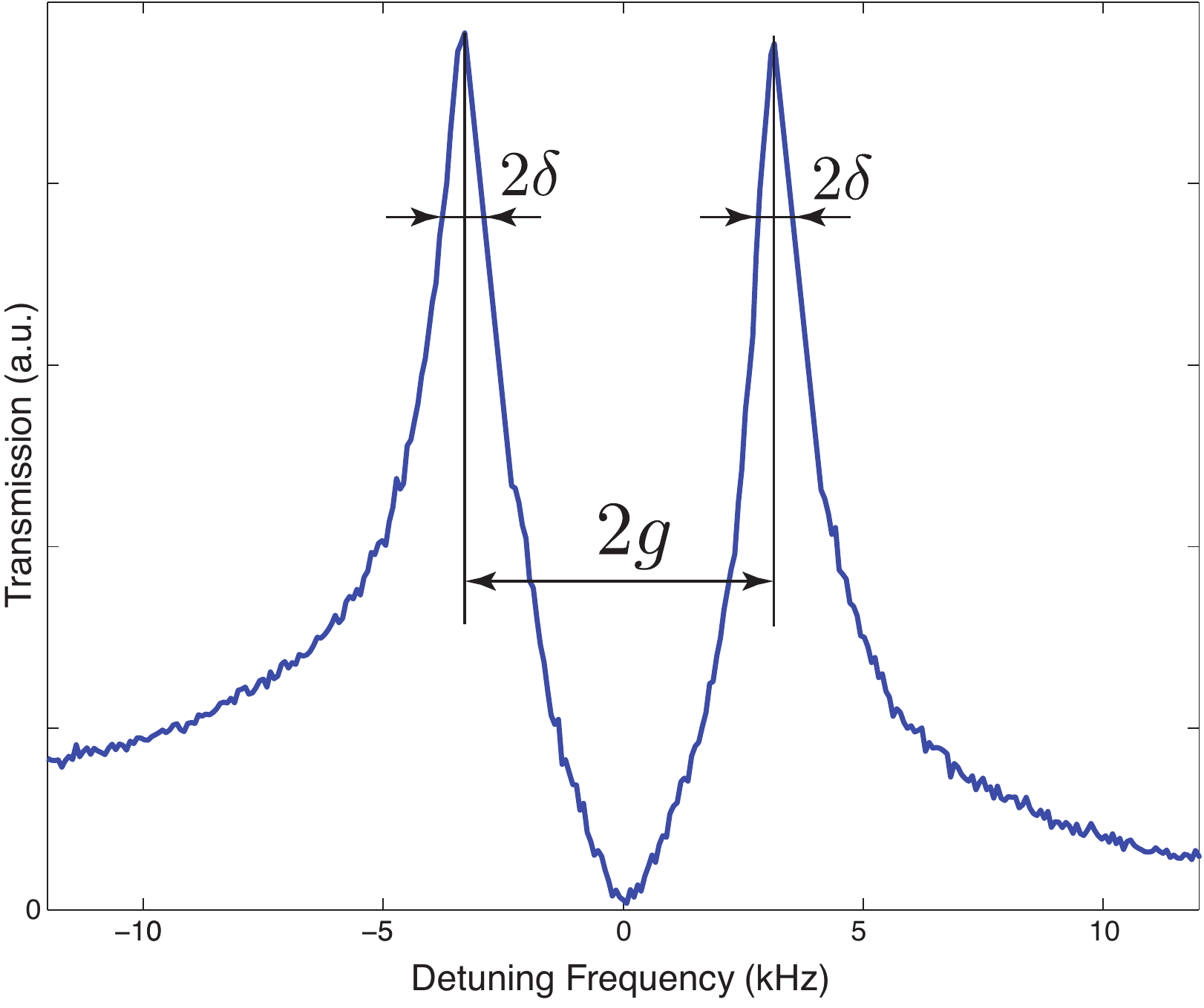}
\caption{\label{cut} (Color online) Transmission through the WGC at \hbox{$B_{\text{ext}}=-0.5$ mT} as a function of detuning frequency from $f_c$. This demonstrates equal losses and the minimal splitting due to the dielectric properties of the material, with $2g=6.46$~kHz and $g/\delta=5.4$}
\end{figure}

Figure~\ref{cut} demonstrates the mode splitting at \hbox{$B_{\text{ext}}=-0.5$~mT}. Both states exhibit equal losses, demonstrating the strong coupling between the photon and its spin angular momentum. This situation is similar to traditional cavity QED experiment where a photon mode is strongly coupled to a specific atomic transition. The strength of the coupling is $2g=6.46$~kHz, and the line width is $2\delta=1206$~Hz, with $g/\delta=5.4$. It should be emphasized that ESR transitions of dilute impurities of the crystal are not explicitly present in the region of interest (shown in Fig.~\ref{cut}). Only the far outside `tail' of the characteristic Lorentzian lineshape of such an ESR is utilized to manipulate the permeability tensor $\hat\mu(B_{\text{ext}})$ of the crystal . This is completely described by the macroscopic properties of the matter, and unlike in traditional QED no microscopic properties of the matter are utilised directly.

\section{Theoretical Description}

The WGC is represented by a uniform, homogeneous, time-invariant, (gyro)-anisotropic medium. The action of the electromagnetic field  in such medium in the absence of charges and currents is given by
\begin{multline}
\label{G001GF}
\displaystyle S = \frac{1}{2}\int_{t_0}^{t_i}\int \left( \mathbf{E}^\dagger (\mathbf{r},t)\cdot \hat\varepsilon \cdot \mathbf{E}(\mathbf{r},t) - \right. \\
\left. \mathbf{H}^\dagger (\mathbf{r},t)\cdot \hat\mu \cdot \mathbf{H}(\mathbf{r},t)\right) d^3\mathbf{r} \text{ }dt,
\end{multline}
where $t$ and $\mathbf{r}$ represent time and the vector displacement from the origin, $\mathbf{E}$ and $\mathbf{H}$ are the intensities of the electric and magnetic fields respectively, and $\hat\varepsilon$ and $\hat\mu$ are the permittivity and permeability tensors.
The corresponding Hamiltonian can be written:
\begin{multline}
\label{G002GF}
\displaystyle H = \frac{1}{2}\int \left( \mathbf{E}^\dagger (\mathbf{r},t)\cdot \hat\varepsilon \cdot \mathbf{E}(\mathbf{r},t)+\right. \\
\left. \mathbf{B}^\dagger (\mathbf{r},t)\cdot (\hat\mu^{-1})^\dagger\cdot \mathbf{B}(\mathbf{r},t)\right) d^3\mathbf{r},
\end{multline}
where $\mathbf{B}$ is the magnetic field, found using the relationship $\hat\mu^{-1}\hat\mu=\hat{I}$ (where $\hat{I}$ is the identity tensor). 

The cavity material properties are given by the two tensors $\hat{\varepsilon}$ and $(\hat\mu^{-1})^\dagger$ in the cavity Hamiltonian (Eq. \ref{G002GF}). These tensors act on a photon spin to change its state, demonstrating operator-like behaviour. The tensors can be represented in the form:  
\begin{align}
\label{G005GF}
\displaystyle \hat{\varepsilon} &= \varepsilon \big(\hat{I} + \hat{\eta}\big),\\
 \displaystyle (\hat\mu^{-1})^\dagger &={\mu}^{-1}\big(\hat{I}+ \hat{\nu}\big),
\end{align}
where $\varepsilon$ and ${\mu}^{-1}$ are scalars representing the average diagonal value of the corresponding quantity. Since any influence of the longitudinal direction is neglected, i.e. no longitudinal component of the field exists, all matrices in Eq. \ref{G005GF} can be represented as square. The last term in both definitions is considered to be a small perturbation of the isotropic components of the parameter due to the anisotropy and gyrotropicity of the medium. 

In a WGC, the electromagnetic wave propagates around the inner surface of a cylindrical cavity. The resonance of such a cavity is set by the number of variations of the field intensity about the azimuthal angle $\phi$. WG cavities exhibit two types of mode with well defined linear polarisation: WGH modes, with dominant $\mathbf{B}$ field along the cylindrical $z$-axis and $\mathbf{E}$ field in the radial direction, and WGE modes with dominant $\mathbf{E}$ field oriented axially, and $\mathbf{B}$ field radially\cite{karim1}.
All modes are hybrid modes to some extent, but the small longitudinal components of the field are ignored here. This approximation is the same as the case of an infinitely long cylindrical cavity, or a mode with very large azimuthal wave number. The assumption reduces the analysis of the system to two transverse dimensions. The two types of waves correspond to the two orthogonal directions of the wave-polarization in the cylindrical geometry, stated further as $\alpha=x=r$ (the cavity radius) and $\alpha=y$ (the cylinder axis). Due to the crystal anisotropy, these two polarizations are not degenerate in frequency and thus the same wavenumber ${k}$ corresponds to two resonance frequencies $\omega_{\text{WGH}}$ and $\omega_{\text{WGE}}$ with a particular linear wave-polarization. In this study, we consider the case of a defined polarisation at a single resonance frequency $\omega_{k\alpha}=\omega_0$. 


For the case of a WGC operating in a particular resonant mode with wavenumber ${k}$ along the azimuthal angle $\phi$, and a defined linear polarisation $\alpha$, the field components in Eq. \ref{G002GF} can be written as follows\cite{allen}:
\begin{align}
\label{G003GF}
\displaystyle \mathbf{E}(\mathbf{r},t) 	&\approx \hat{E}_\alpha(\phi,t) \nonumber \\
															&= i\sqrt{\frac{2\pi\hbar}{V\varepsilon}}\sqrt{\omega_{k\alpha}}\Big[\hat\sigma_\alpha \hat{a} e^{ik\phi}-\hat\sigma_\alpha^\dagger \hat{a}^\dagger e^{-ik\phi}\Big],\\
\displaystyle \mathbf{B}(\mathbf{r},t) &\approx \hat{B}_\alpha(\phi,t) \nonumber\\
															&= \pm i\sqrt{\frac{2\pi\hbar}{V\varepsilon}}\frac{ k}{\sqrt{\omega_{k\alpha}}}\Big[\hat\sigma_\beta \hat{a} e^{ik\phi}-\hat\sigma_\beta^\dagger \hat{a}^\dagger e^{-ik\phi}\Big]
\end{align}
where $\hat\sigma_\alpha$ and $\hat\sigma_\beta$ are polarisation vector operators along $x$ or $y$, and $a^\dagger$ \& $a$ are the creation and annihilation operators for the cavity photon. Note that only one linearly polarised wave exists at a given angular frequency.  

Linearly polarised modes of the WGC can be described using a basis vector of two helicities $\hat\sigma_R$ and $\hat\sigma_L$, i.e. right and left circular polarizations\cite{allen}:
\begin{align}
\label{G004GF}
\displaystyle \hat\sigma_x &= \frac{1}{\sqrt{2}}(\hat\sigma_R+\hat\sigma_L)=\frac{1}{\sqrt{2}}(\Ket{R}\Bra{L}+\Ket{L}\Bra{R}), \text{ or }\\
 \displaystyle \hat\sigma_y &= \frac{1}{i\sqrt{2}}(\hat\sigma_L-\hat\sigma_R)=\frac{1}{i\sqrt{2}}(\Ket{L}\Bra{R}-\Ket{R}\Bra{L})
\end{align}
 These two circular polarizations represent the spin of a photon. Using the language of spin-$\frac{1}{2}$ particles, a photon has spin 1 if it is in the $\Ket{R}$ state (right-polarized) and -1 if it is in the  $\Ket{L}$ state (left-polarized), which are the eigenstates of the photon spin angular momentum. From this definition, the operators $\hat\sigma_R$ and $\hat\sigma_L$ reverse the photon helicity. This situation is similar to the $\hat\sigma_+=\Ket{e}\Bra{g}$ and $\hat\sigma_-=\Ket{g}\Bra{e}$ operators for a spin-$\frac{1}{2}$ particle, where these operators introduce or remove an excitation from the system and change the atomic spin angular momentum by $\pm1$. Other important properties of spin-$\frac{1}{2}$ particles, such as the fact that $\hat\sigma_\alpha = (\hat\sigma_\alpha)^\dagger$ and $\hat\sigma_R=(\hat\sigma_L)^\dagger$ also hold true in the case of photon spin. In the case of photon spin, the spin operator acting in the $z$-direction can be defined as $\hat\sigma_z=\Ket{R}\Bra{R}-\Ket{L}\Bra{L}$.

The action on the photon spin angular momentum of the first terms of the sums in the material tensors (Eq. \ref{G005GF}) is trivial, i.e. the associated identity matrices do not change the spin state of the photon. The second terms of the sums give a non-trivial change of the state that can be represented generically in terms of the following matrices:
\begin{equation}
	\label{G006GF}
\hat{\eta} =\left( \begin{array}{llll}
	\Bra{R}\hat{\eta}\Ket{R} & \Bra{R}\hat{\eta}\Ket{L}   \\
	\Bra{L}\hat{\eta}\Ket{R} & \Bra{L}\hat{\eta}\Ket{L}  \\
 \end{array} \right) \\
\end{equation}
\begin{equation}
	\label{G006GF-2}
 \hat{\nu} =\left( \begin{array}{llll}
	\Bra{R}\hat{\nu}\Ket{R} & \Bra{R}\hat{\nu}\Ket{L}   \\
	\Bra{L}\hat{\nu}\Ket{R} & \Bra{L}\hat{\nu}\Ket{L}  \\
 \end{array} \right),
\end{equation}
where for a gyrotropic medium it is known that $\hat{\eta}=\hat{\eta}^\dagger$ and $\hat{\nu}=\hat{\nu}^\dagger$. In addition, $\eta_{11}=-\eta_{22}$ and $\nu_{11}=-\nu_{22}$ due to the choice of scalars $\varepsilon$ and $\mu^{-1}$ as averages of the corresponding properties between the two dimensions. If the material is isotropic then the diagonal elements in both matrices $\hat{\nu}$ and $\hat{\eta}$ vanish. However, in the case of an isotropic non-gyrotropic medium, both matrices become zero. The matrix representations (Eq. \ref{G006GF} and \ref{G006GF-2}) are due to the law of spin angular momentum conservation in a matter-field interaction where $\Ket{R}$ and $\Ket{L}$ are the eigenstates of the field.

Such a representation of the medium, together with the photon polarisation decomposition into a circular polarisation basis, (\ref{G004GF}) suggests the following:
\begin{align}
\label{G007GF}
\displaystyle H_\alpha & =  \frac{1}{4}\hbar\omega_0\big[\hat\sigma_\alpha\hat\sigma_\alpha+\hat\sigma_\beta\hat\sigma_\beta+ \hat\sigma_\alpha\hat{\eta}\hat\sigma_\alpha+\hat\sigma_\beta\hat{\nu}\hat\sigma_\beta\big]\big[a^\dagger a+a a^\dagger \big]\nonumber\\
\displaystyle & ={}\hbar\omega_0\big[1+ \frac{1}{2}\hat\sigma_\alpha\hat{\eta}\hat\sigma_\alpha+\frac{1}{2}\hat\sigma_\beta\hat{\nu}\hat\sigma_\beta\big]\Big[a^\dagger a+\frac{1}{2} \Big]
\end{align}
where the $\sigma_\alpha\hat\sigma_\alpha^\dagger=1$ term is due to taking into account the isotropic component and dispersion relationship \hbox{$\omega^2=\frac{k^2}{\mu\varepsilon}$}.
The Hamiltonian (Eq. \ref{G007GF}) clearly shows that the energy of a bosonic mode of a photon depends on its spin state. 
Expanding the second term of the first square bracket, the Hamiltonians of the WGE and WGH cavity modes can be written as
\begin{equation}
\label{G010GF}
\begin{split}
H_x  & = {}\hbar\omega_0\Big(1+ \frac{1}{2}\hat\sigma_x\hat{\eta}\hat\sigma_x+\frac{1}{2}\hat\sigma_y\hat{\nu}\hat\sigma_y\Big)\Big[a^\dagger a+\frac{1}{2} \Big]\\
& =\hbar\omega_0\left(1+ \frac{\eta_{22}+\nu_{22}}{2}\hat\sigma_z+\Im\frac{\eta_{21}-\nu_{21}}{\sqrt{2}}\hat\sigma_y \right.\\
 &\hspace{2cm} +\left.\Re\frac{\eta_{21}-\nu_{21}}{\sqrt{2}}\hat\sigma_x\right)\Big[a^\dagger a+\frac{1}{2} \Big]
\end{split}
\end{equation}
and 
\begin{equation}
\label{G011GF}
\begin{split}
\displaystyle H_y  &= \hbar\omega_0\Big(1+ \frac{1}{2}\hat\sigma_y\hat{\eta}\hat\sigma_y+\frac{1}{2}\hat\sigma_x\hat{\nu}\hat\sigma_x\Big)\Big[a^\dagger a+\frac{1}{2} \Big]\\
\displaystyle &={}\hbar\omega_0\Big(1+ \frac{\eta_{22}+\nu_{22}}{2}\hat\sigma_z+\Im\frac{\nu_{21}-\eta_{21}}{\sqrt{2}}\hat\sigma_y\\
&\hspace{2cm}+\Re\frac{\nu_{21}-\eta_{21}}{\sqrt{2}}\hat\sigma_x\Big)\Big[a^\dagger a+\frac{1}{2} \Big].
\end{split}
\end{equation}

For both the WGE and WGH polarizations of the mode, the results (\ref{G010GF}) and (\ref{G011GF}) demonstrate a dependence of the angular frequency of the bosonic mode of the photon on its spin angular momentum state. This dependence vanishes if the medium is isotropic ($\eta_{22}+\nu_{22}=0$) and not gyrotropic ($\Im\{\nu_{21}-\eta_{21}\}=0$ and $\Re\{\nu_{21}-\eta_{21}\}=0$) leaving the usual form of a single harmonic oscillator (HO). 
Otherwise, splitting of the mode spectrum can be observed due to the states $\Ket{R}$ and $\Ket{L}$ of the TLS emerging from the photon spin angular momentum.
So, strong coupling of the photon spin and bosonic degrees of freedom is possible. Experimentally, this is observed as two realisations of the bosonic mode for two states of the photon spin $\Ket{R}$ or $\Ket{L}$. Which photon spin state is higher in energy (the exited state) is set by the material parameters. Thus for an anisotropic, weakly gyrotropic ($|\Im\{\nu_{21}-\eta_{21}\}|\ll|\eta_{22}+\nu_{22}|$) medium, if $\eta_{22}+\nu_{22}>0$ then the $\Ket{R}$ state is the excited state, and $\Ket{L}$ is the ground state. This is the situation we assume in the present work. 
The interesting feature of the interaction we suggest is that the ensemble of TLS cannot be saturated unless the material parameters themselves are non-linear, otherwise the number of TLS is always equal to the number of bosons in the mode. This fact manifests an important difference between the present system and traditional cavity QED involving an ensemble of two level systems. Whereas the latter usually exhibits nonlinear effects when the ensemble is close to saturation, the former is always linear and representable by a system of two coupled HOs. However, the TLS ensemble case can also be approximated by the set of HOs in the case of weak excitation\cite{ritsch}.



\section*{Conclusion}

In this work, we have demonstrated similarities between traditional QED\cite{vernooy}, which employs an interaction between quantised electromagnetic modes with a two-level state system (a spin-$\frac{1}{2}$ particle), and the approach we propose, which employs interaction between bosonic behaviour of a photon in a cavity and the photon's own spin. Most of the similarities are due to the fact that, although a photon is a spin-1 particle carrying spin $\pm1$, it has only two states of spin angular momentum because of its zero rest mass. These two eigenstates imply that a photon can, in principle, behave as a qubit like other quantum TLSs based on eigenstates of a photon spin angular momentum operator\cite{hiroshi,karassiov}: $\Ket{\Psi}=\alpha\Ket{R}+\beta\Ket{L}$, where the complex coefficients are subject to $\alpha^2+\beta^2=1$. However, unlike typical Jaynes-Cummings systems, the spin-phonon system analysed in this work is always linear and can thus always be described as a system of two coupled HOs. This fact prevents its direct utilization as a qubit without the introduction of an additional nonlinearity, for instance through the spins controlling the interaction, a superconducting nonlinear junction, or a nonlinear measurement scheme.

Nevertheless, the narrow linewidths of WG modes allow these cavities to achieve operation in the strong coupling regime where a photon loses its identity in terms of its own spin. In this case, unlike in other work\cite{mazzei}, matter is treated collectively by the tensor macroscopic description, which breaks the reflection symmetry leading to energy splitting between photon spin states. This approach to matter-field interaction is significantly different from the traditional one extensively implying microscopic quantum phenomena in the material. It is also different from the experiment utilising quantum regimes of cavity photons\cite{cqed1,cqed2} since only bosonic degrees of freedom are utilised in this approach. This work points out another source of quantum states, one originating in the photon spin degree of freedom.  So, this work demonstrates possibility of achieving high $Q$-factor values in cavities with broken symmetry, and finally, we demonstrate an effective method of controlling this phenomenon through natural dilute impurities with the application of an external magnetic field.





\section*{Acknowledgements}

This work was supported by the Australian Research Council Grant No. CE110001013 and FL0992016.

\section*{References}


%

\end{document}